\documentclass[journal]{IEEEtran}

\usepackage{graphicx}  

\begin{document}
%
\title{A Study of Multiscale Density Fluctuation Measurements}
%
%
\author{Nils~P.~Basse,~\IEEEmembership{Member,~IEEE}
\thanks{Manuscript submitted August 23, 2007. This work
was supported by the U.S. Department of Energy, Office of Fusion
Energy Sciences.}
\thanks{N. P. Basse was with the Plasma Science and Fusion
Center, Massachusetts Institute of Technology, Cambridge, MA-02139,
USA. He is now with ABB Switzerland Ltd., Corporate Research,
Segelhofstrasse 1, CH-5405 Baden-D\"attwil, Switzerland (e-mail:
nils.basse@ch.abb.com).}}

\markboth{IEEE TRANSACTIONS ON PLASMA SCIENCE}{BASSE: A STUDY OF
MULTISCALE DENSITY FLUCTUATION MEASUREMENTS}
%



\maketitle

\begin{abstract}
Intriguing parallels between density fluctuation power versus
wavenumber on small (mm) and large (Mpc) scales are presented. The
comparative study is carried out between fusion plasma measurements
and cosmological data. Based on predictions from classical fluid
turbulence theory, we argue that our observations are consistent
with 2D turbulence. The similar dependencies of density fluctuations
on these disparate scales might indicate that primordial turbulence
has been expanded to cosmological proportions.
\end{abstract}


%
\IEEEpeerreviewmaketitle

\section{Introduction}
\label{sec:intro}

It is a very human trait to compare new observations to previous
experience. Our chance encounter with measurements of the spectral
power of density fluctuations on Mpc scales lead us to the
conclusion that corresponding mm scale measurements in fusion
plasmas have surprisingly similar features \cite{basse1}. We are of
the opinion that this correspondence could have a significant impact
on current ideas regarding the formation of the universe.

Let us briefly present our reasoning: Fusion plasmas are turbulent,
whereas density fluctuations on cosmological scales are not.
However, the cosmological fluctuations might be what has been dubbed
"fossilized turbulence" \cite{gamow,gibson1}, i.e. static images of
primordial turbulence. This original hot big bang turbulence is in
our picture represented by fusion plasma turbulence. So the emerging
understanding is as follows: (i) turbulence was generated before the
inflationary expansion of the universe, (ii) as the universe cooled
and expanded, the primordial turbulence fossilized and is visible on
cosmological scales today. The theoretical basis of this hypothesis
is outlined in Refs. \cite{gibson2,gibson3}.

We show in this paper that both sets of measurements fit the shape
expected from 2D fluid turbulence theory. According to our
interpretation, this implies that early turbulence was 2D.

The fusion plasma measurements presented in this paper are of
fluctuations in the electron density. Phase-contrast imaging (PCI)
\cite{mazurenko} is being used in the Alcator C-Mod tokamak
\cite{hutch} and small-angle collective scattering (SACS)
\cite{saffman} was used in the Wendelstein 7-AS (W7-AS) stellarator
\cite{renner}.

We specifically study density fluctuation power $P$ versus
wavenumber $k$ (also known as the wavenumber spectrum) in C-Mod and
W7-AS. These wavenumber spectra characterize the nonlinear
interaction between turbulent modes having different length scales.
Our explicit assumption is that turbulence in stellarators and
tokamaks is comparable.

The second part of our measurements, a cosmological wavenumber
spectrum constructed from a variety of sources, has been published
in Ref. \cite{tegmark1} and was subsequently made available to us
\cite{tegmark2}. The measurements were used to constrain
cosmological variables, e.g. the matter density $\Omega_m$ and
neutrino masses - for further details see Refs.
\cite{tegmark1,tegmark3}.

The paper is organized as follows: In Sec. \ref{sec:wano} we analyze
fusion plasma and cosmological wavenumber spectra. Thereafter we
treat the dimensionality of the measurements in Sec. \ref{sec:dim}.
We discuss the hot big bang turbulence theory in Sec. \ref{sec:bbt}
and conclude in Sec. \ref{sec:conc}.

\section{Wavenumber spectra}
\label{sec:wano}

\begin{figure*}
\centering
\includegraphics[height=8.5cm]{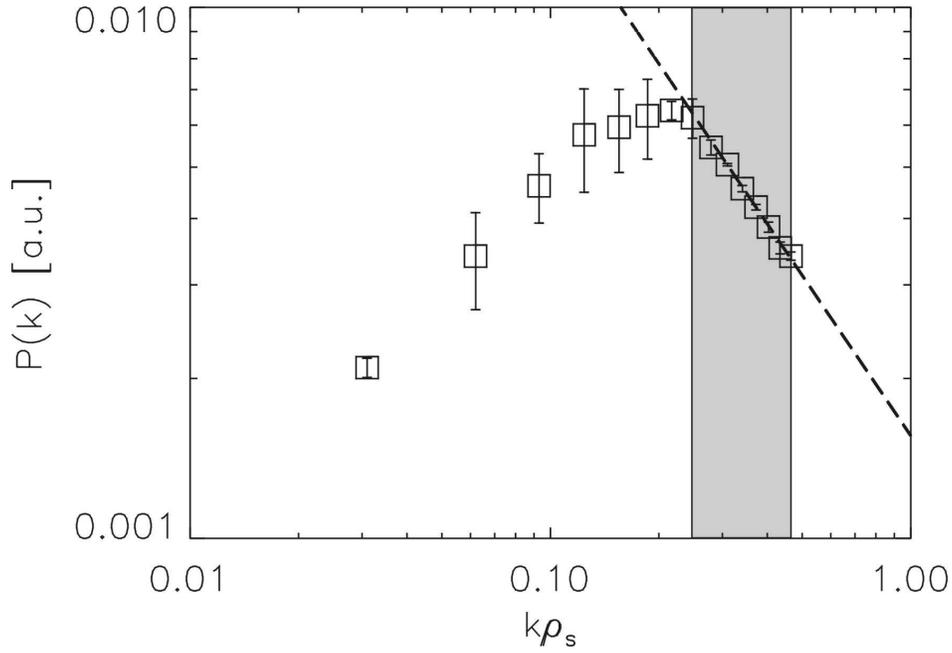}
\caption{\label{fig:pci} Wavenumber spectrum of broadband turbulence
in C-Mod. Squares are measured points. The dashed line is a fit to
(\ref{eq:pow_decay}); the semi-transparent rectangle indicates which
points are included to make the fit. The measurements are taken from
Fig. 11 in Ref. \cite{basse2}.}
\end{figure*}

We begin by studying the fusion plasma wavenumber spectrum shown in
Fig. \ref{fig:pci}. The plot shows PCI measurements along with a fit
to

\begin{equation}
P(k\rho_s) \propto (k\rho_s)^{-m}, \label{eq:pow_decay}
\end{equation}

\noindent where $\rho_s$ is the ion Larmor radius at the electron
temperature and $m$ is a constant. The measurements were made in a
low confinement mode C-Mod plasma, see Fig. 11 in Ref.
\cite{basse2}. The wavenumbers measured have been multiplied by
$\rho_s$, which for this case is 0.6 mm. This is the value at 80 \%
of the plasma radius where the electron temperature is 400 eV, the
toroidal magnetic field is 6.4 T and the working gas is Deuterium.

Our fit to the indicated PCI data yields $m$ = 1.0 $\pm$ 0.03.

All fits shown in this paper have a normalized $\chi^2$ $\le$ 1,
ensuring a satisfactory quality. The error bars are standard
deviations and the semi-transparent rectangles indicate which points
are included to make the fits.

\begin{figure*}
\centering
\includegraphics[height=8.5cm]{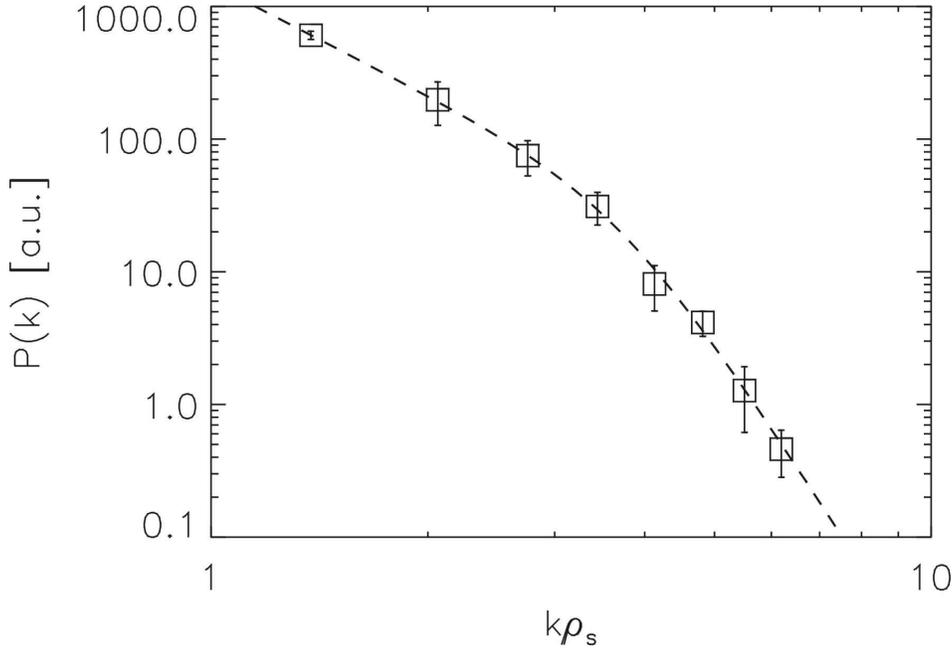}
\caption{\label{fig:sacs} Wavenumber spectrum of broadband
turbulence in W7-AS. Squares are measured points. The dashed line is
a fit to (\ref{eq:nabhan}); all points are included to make the fit.
The measurements are taken from Fig. 12 in Ref. \cite{basse3}.}
\end{figure*}

In Fig. \ref{fig:sacs} we show SACS measurements at somewhat larger
wavenumbers compared to the PCI data. Again, the measured
wavenumbers have been multiplied by $\rho_s$, which in this case is
1 mm. This value is also at 80 \% of the plasma radius where the
electron temperature is 300 eV, the toroidal magnetic field is 2.5 T
and the working gas is Hydrogen.

The SACS measurements are fitted to

\begin{equation}
P(k\rho_s) \propto \frac{(k\rho_s)^{-p}}{1 +
(k\rho_s/(k\rho_s)_0)^q}, \label{eq:nabhan}
\end{equation}

\noindent where $p$ = 2.8 $\pm$ 0.6 and $q$ = 5.7 $\pm$ 1.3 are
constants. The functional form in (\ref{eq:nabhan}) is taken from
Ref. \cite{nabhan}. Basically this equation describes two
power-laws, where $P \propto (k\rho_s)^{-p} = (k\rho_s)^{-2.8}$ for
medium wavenumbers and $P \propto (k\rho_s)^{-p-q} =
(k\rho_s)^{-8.5}$ for large wavenumbers. The transitional
$(k\rho_s)_0$ is in our case 3.7. The W7-AS data have been taken
from Fig. 12 in Ref. \cite{basse3}.

It is at this point relevant to note that the medium wavenumber
fusion plasma exponent is not always three (or 2.8), it typically
varies between three and four depending on specific plasma
conditions \cite{honore,zoletnik,hennequin}. Presumably this is due
to different instabilities driving turbulence for varying operating
conditions, leading to forcing centered at changing scales.

\begin{figure*}
\centering
\includegraphics[height=8.5cm]{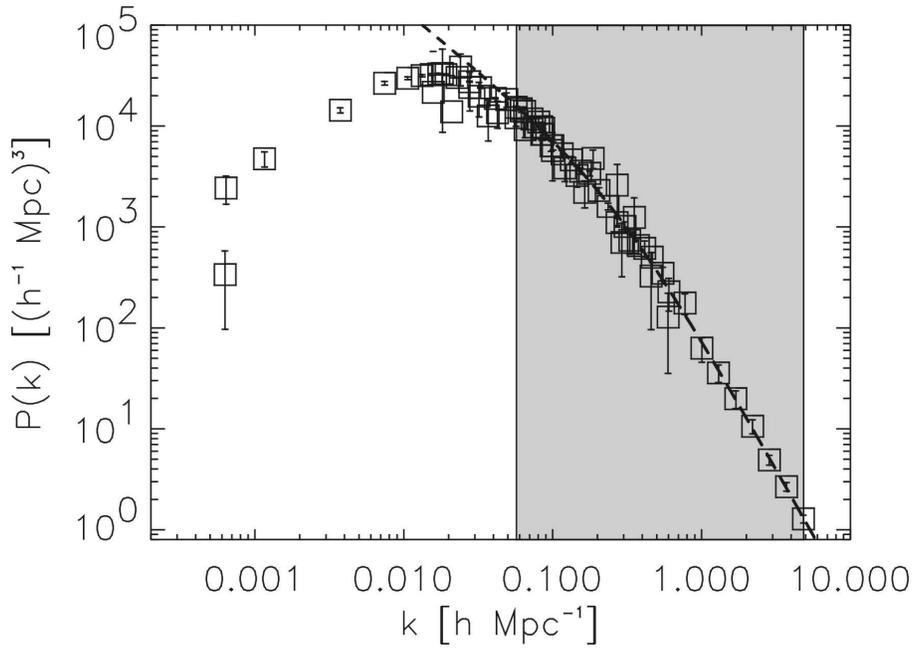}
\caption{\label{fig:cosmo} Wavenumber spectrum of the combined
cosmological measurements. Squares are measured points. The dashed
line is a fit to (\ref{eq:nabhan}); the semi-transparent rectangle
indicates which points are included to make the fit. The
measurements are taken from Fig. 38 in Ref. \cite{tegmark1}.}
\end{figure*}

The cosmological wavenumber spectrum is shown in Fig.
\ref{fig:cosmo}. The measurements are fitted to (\ref{eq:nabhan}),
but using $k$ instead of $k\rho_s$; in this case, $p$ = 1.2 $\pm$
0.1 and $q$ = 1.4 $\pm$ 0.05 are constants. Here, $P \propto k^{-p}
= k^{-1.2}$ for small wavenumbers and $P \propto k^{-p-q} =
k^{-2.6}$ for medium wavenumbers. The transitional wavenumber $k_0$
is 0.3 h Mpc$^{-1}$. Here, $h = H_0$/(100 km/s/Mpc) $\simeq$ 0.7,
where $H_0$ is the Hubble parameter observed today.

\section{Dimensionality of the measured fluctuations}
\label{sec:dim}

We begin Sec. \ref{sec:dim} by summarizing our findings on the
dependencies of power on wavenumber in Sec. \ref{sec:wano}:

\begin{eqnarray}
{\bf Small \ wavenumbers:} \nonumber\\ P(k) \propto k^{-1.0} {\rm
(fusion) \ or \ } P(k)
\propto k^{-1.2} {\rm (cosmology).} \nonumber\\
{\bf Medium \ wavenumbers:} \nonumber\\ P(k) \propto k^{-2.8} {\rm
(fusion) \ or \ } P(k)
\propto k^{-2.6} {\rm (cosmology)}. \nonumber\\
{\bf Large \ wavenumbers:} \nonumber\\ P(k) \propto k^{-8.5} {\rm
(fusion).} \label{eq:power_rules}
\end{eqnarray}

Our measured density fluctuation power is equivalent to the
$d$-dimensional energy spectrum $F_d(k)$
\cite{tennekes,frisch,antar}

\begin{eqnarray}
P(k) = F_d(k) = \frac{E(k)}{A_d} \nonumber \\ \nonumber \\ A_1 = 2
\hspace{2cm} A_2 = 2\pi k \hspace{2cm} A_3 = 4\pi k^2,
\label{eq:e_spec}
\end{eqnarray}

\noindent where $A_d$ is the surface area of a sphere having radius
$k$ and dimension $d$.

We can convert our results in (\ref{eq:power_rules}) either under
the 2D turbulence assumption:

\begin{eqnarray}
{\bf Small \ wavenumbers:} \nonumber\\ E(k) \propto k^{0.0} {\rm
(fusion) \ or \ } E(k)
\propto k^{-0.2} {\rm (cosmology).} \nonumber\\
{\bf Medium \ wavenumbers:} \nonumber\\ E(k) \propto k^{-1.8} {\rm
(fusion) \ or \ } E(k)
\propto k^{-1.6} {\rm (cosmology)}. \nonumber\\
{\bf Large \ wavenumbers:} \nonumber\\ E(k) \propto k^{-7.5} {\rm
(fusion)}. \label{eq:2d}
\end{eqnarray}

\noindent or under the 3D turbulence assumption:

\begin{eqnarray}
{\bf Small \ wavenumbers:} \nonumber\\ E(k) \propto k^{1.0} {\rm
(fusion) \ or \ } E(k)
\propto k^{0.8} {\rm (cosmology).} \nonumber\\
{\bf Medium \ wavenumbers:} \nonumber\\ E(k) \propto k^{-0.8} {\rm
(fusion) \ or \ } E(k)
\propto k^{-0.6} {\rm (cosmology)}. \nonumber\\
{\bf Large \ wavenumbers:} \nonumber\\ E(k) \propto k^{-6.5} {\rm
(fusion)}. \label{eq:3d}
\end{eqnarray}

The established picture of 2D fluid turbulence is: (i) turbulence is
forced on an intermediate scale $k_{\rm f(2D)}$, (ii) energy is
transferred to larger scales by the inverse energy cascade, $E(k)
\propto k^{-5/3}$ \cite{chen2}, and enstrophy is transferred to
smaller scales by the forward enstrophy cascade, $E(k) \propto
k^{-3}$ \cite{chen3}, and (iii) enstrophy is dissipated at the
smallest scales \cite{chen1}.

For 3D turbulence the following process occurs: (i) turbulence is
forced on a large scale $k_{\rm f(3D)}$, (ii) energy is transferred
to smaller scales by the forward energy cascade, $E(k) \propto
k^{-5/3}$, and (iii) energy is dissipated at the smallest scales.

It is interesting to note that in Ref. \cite{neumann}, two
dependencies of the 3D energy spectrum on wavenumber in the
dissipation range are considered: One is an exponential falloff, the
other claims that $E(k) \propto k^{-7}$ and was proposed by W.
Heisenberg. This power-law is quite close to the one we found for
fusion plasmas at large wavenumbers.

To determine whether 2D or 3D turbulence is observed, we consider
the power-laws for medium wavenumbers: The exponents should roughly
be in the range [-3, -5/3] for 2D turbulence and about -5/3 for 3D
turbulence. Equations (\ref{eq:2d}) and (\ref{eq:3d}) indicate that
the observed 2D slopes are close to the expected power-laws and that
the 3D slopes are too shallow. We note that the 2D slopes are closer
to the value for the inverse energy cascade than for the forward
enstrophy cascade. The reason for this is not understood.

Turbulence in fusion plasmas is approximately 2D, since transport
along magnetic field lines is nearly instantaneous. For this reason,
fluctuations are measured parallel to the major radius of the
machine, i.e. perpendicular to the confining magnetic field.

The reason we chose to analyze fusion plasma data was simply a
matter of having available measurements and expertise in that field.
Any turbulent 2D plasma should display similar characteristics.

\section{Hot big bang turbulence} \label{sec:bbt}

In Ref. \cite{basse1} we suggested that the observed plasma
turbulence might originate during an early phase in the formation of
the universe. Recent theoretical work on the role of hot big bang
turbulence in the primordial universe \cite{gibson3} lends support
to this assumption:

In this theory, turbulence observed today was created before
cosmological inflation by inertial-vortex forces leading to an
inverse big-bang turbulence cascade with a -5/3 power-law exponent.
Turbulence in the plasma epoch has low Reynold's numbers $\sim$
10$^2$ according to this picture and the preceding quark-gluon
plasma has a large gluon viscosity that acts to damp the big bang
turbulence. The claim is that hot big bang temperature turbulence is
fossilized before the universe cools to the Grand Unified Theory
strong force freeze-out temperature 10$^{28}$ K. Later, during
nucleosynthesis, fossil temperature turbulence was converted to
fossil turbulence patterns in e.g. density turbulence
\cite{gibson2}.

Analysis of power spectra of cosmic microwave background radiation
temperature anisotropies shows that they most likely have a
turbulent origin, supporting the idea of turbulence generated in the
big bang or the plasma epoch \cite{bershadskii}.

\section{Conclusions}
\label{sec:conc}

The fact that density fluctuations on small (fusion plasma) and
large (cosmological) scales can be described by similar functional
dependencies, approximately consistent with 2D fluid turbulence,
might indicate that (plasma) turbulence at early times has been
fossilized and expanded to cosmological proportions.

Our conjecture concerning the primordial turbulence reflected in our
wavenumber spectra can be described as follows: Forcing occurs at an
intermediate scale $k_{\rm f(2D)}$. The inverse energy cascade leads
to spectral condensation at large scales and the forward enstrophy
cascade leads to enstrophy transfer towards smaller scales. At very
small scales, enstrophy is dissipated.

Our interpretation of the measured wavenumber spectra is consistent
with the theoretical framework on hot big bang turbulence presented
in Ref. \cite{gibson3} and references therein.

\section*{Acknowledgment}
This work was supported at MIT by the Department of Energy,
Cooperative Grant No. DE-FC02-99ER54512. We thank M. Tegmark for
providing the cosmological measurements analyzed in this paper.

\begin{biography}[{\includegraphics[width=1in,height=1.25in,clip,keepaspectratio]{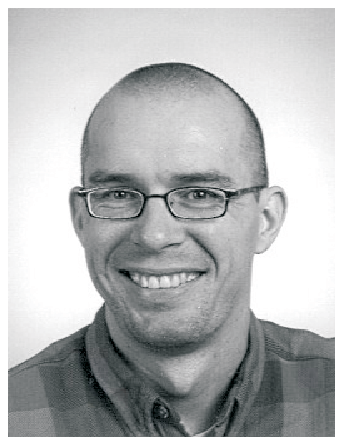}}]{Nils P. Basse} received the B.Sc., M.Sc., and Ph.D.
degrees from the Niels Bohr Institute, University of Copenhagen,
Copenhagen, Denmark, in 1996, 1998, and 2002, respectively. He is a
Scientist with ABB Switzerland Ltd., Corporate Research. He was a
Postdoctoral Associate at the Plasma Science and Fusion Center,
Massachusetts Institute of Technology (MIT), Cambridge, from 2002 to
2005. His present research interests include plasmas in medium- and
high-voltage circuit breakers.
\end{biography}




\end{document}